\documentstyle[prl,aps,epsfig,multicol]{revtex}

\begin{document}

\large


\newcommand{\om}{\omega}

\begin{center}

{\bf Nuclear Reaction Rates in Dense Plasmas}

V.~I.~Savchenko

{\it $^+$ Princeton University, PPPL, Forrestal Campus, Princeton, 
N. J. 08543}

\end{center}

\begin{abstract}
We solve the quasiclassical problem of tunneling through an external
potential barrier in a dense plasma, where the tunneling
particles undergo 
simultaneous collisions with other particles in thermodynamic
equilibrium. Under such conditions 
the spectral density of states available to the particle has a
Lorentz shape, rather than the delta function, which leads to a quantum
tail in the particle momentum distribution function.
We show that, this tail indeed significantly alters the
average nuclear reaction rates, which supports earlier suggestions.
This rate can be many orders of magnitude higher
than would be normally calculated by averaging over the Maxwell
distribution of energies.
\end{abstract}

\begin{multicols}{2}
\dimen100=\columnwidth \setlength{\columnwidth}{3.375in}

The particle distribution over momenta, $f({\bf p})$, acquires
non-Maxwellian tail in 
thermodynamic equilibrium~\cite{tails-galitski} due to quantum
effects, while the 
particle distribution over energies, $\tilde{f}(\om)$, remains Maxwellian. An interesting
suggestion was made in Ref.~\cite{tails-cross-section-star}, that the
rates of various processes, including nuclear reactions, may be
significantly increased by the quantum tail of the {\em momentum}
distribution. A final formula for the ionization rate in a closed form
was obtained in \cite{tails-cross-section-star}, by using
the Born approximation. The formula for the nuclear reaction rates was found
in \cite{neutrino-vlad} under assumption that it is correct to
substitute the particle momentum in the form 
$\epsilon_{\bf p}=p^2/2m$ into the known quasiclassical cross-section
$\sigma(\om)$ and then average it over the particle distribution
over momenta $f({\bf p})$ rather than the distribution over energies
$\tilde{f}(\om)$. This
procedure becomes even more unclear, if we take into account the fact, that the
momentum, ${\bf p}$, and energy, $\om$, of a {\em colliding}
particle are {\em independent} variables 
not connected by the usual dispersion relation $\delta(\om -
{\bf p}^2/2m)$~\cite{tails-galitski,statistical-mechanics-kadanov,tails-cross-section-star}.
Note that the averaging of $\sigma(\om)$ over the distribution
over energies $\tilde{f}(\om)$, which is Maxwellian would
give the usual reaction rates~\cite{rate-gamow} rather than the rates found in
\cite{neutrino-vlad}, which are accelerated by many orders of
magnitude in certain regimes. It is therefore important to find the
nuclear reaction rates from first principles.

In this paper we rigorously
solve the quasiclassical problem of tunneling through the potential
barrier, when the tunneling particles undergo simultaneous collisions
with other particles of the plasma maintained in thermodynamic
equilibrium, (see Fig.~1). The plasma has density and temperature $n_0,
T$ and is fully ionized. We use the Green--function 
technique~\cite{landau-10,diagramms-keldish,diagramms-korenman}, and do
not rely on any assumptions~\cite{neutrino-vlad} about
any averaging procedure. 
We show that, if the requirements 
of the quasiclassical approximation are fulfilled with respect to both the
barrier and collisions, our final result for the nuclear reaction rate
supports the idea of averaging over momentum postulated
in~\cite{neutrino-vlad}. 
\begin{figure}[!h]
         \centering
         \mbox{\epsfig{figure=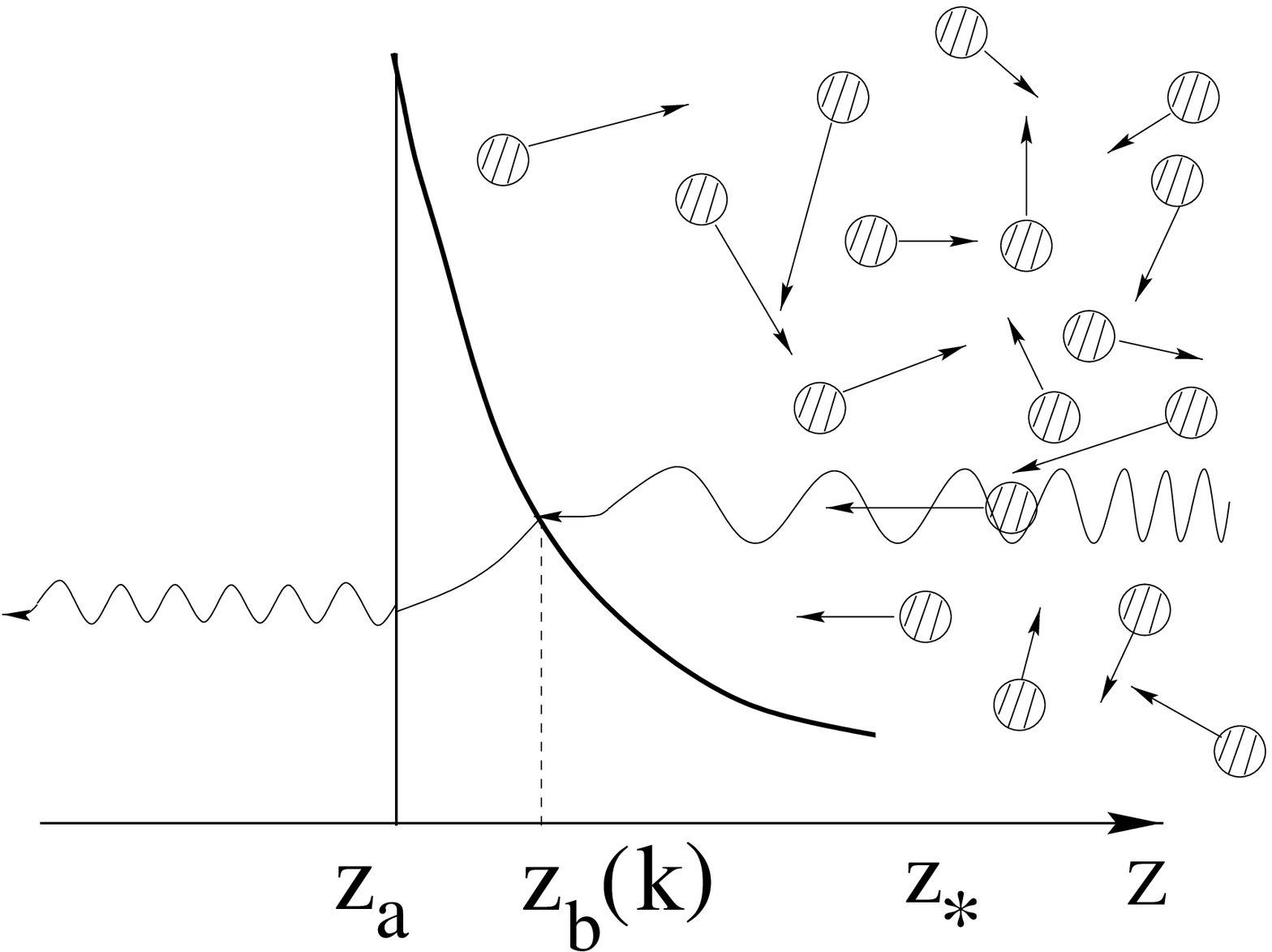, width=2.8
         true in}} 
         \caption{} \label{barrier}
\end{figure}
According to the diagram technique of Keldsyh and
Korenman~\cite{diagramms-keldish,diagramms-korenman} the kinetic
properties of the particles are described by the Green function
$G^{-+}(X_1,X_2)$; we adopt the notation $X=(t,{\bf
x})$. Kinetic equations for this 
function~\cite{diagramms-keldish,diagramms-korenman} include the
interaction between the particles as well as their interaction with
the external field. Therefore, the rate of change of this
function, evaluated at any point after the barrier, will give the
tunneling rate $K({\bf x}, t)$ for the particles colliding
simultaneously with other particles:
\begin{eqnarray*}
K({\bf x}, t)=\left[(\partial_{t_1} - \partial_{t_2}) G^{-+}(t_1 {\bf
x_1}; t_2 {\bf x_2})\right]_{X_1 \rightarrow
X_2=X}&&
\end{eqnarray*}
\begin{equation}
=2\left[\partial_{t_1} G^{-+}(t_1 {\bf 
x_1}; t_2 {\bf x_2})\right]_{X_1 \rightarrow X_2=X}, \label{rate-general}
\end{equation}
In the second line of (\ref{rate-general}) we anticipated that,
$G^{-+}$ depends only on $t_1-t_2$.

Now we will proceed with solution of the kinetic equations for
$G^{-+}(X_1,X_2)$. They can be written as Dyson equations for
$G^{\alpha \beta}_{12}$~\cite{landau-10}:
\begin{equation}
G^{\alpha \beta}_{12}=G^{(0)\alpha \beta}_{12} + \int G^{(0)\alpha
\gamma}_{14} \Sigma^{\gamma \delta}_{43} G^{\delta \beta}_{32} d^4 X_4
d^4 X_3, \label{kinetic-general}
\end{equation}
where we use subscripts $1..4$ to denote dependence on $X_1..X_4$. 
Dependence on $G^{(0)\alpha \beta}_{12}$ can be
eliminated~\cite{landau-10} by acting 
on both sides of the Eq.~(\ref{kinetic-general}) with the operator
\begin{eqnarray*}
&&\hat{G}^{-1}_1=i\frac{\partial}{\partial t_1} + \frac{\Delta_1}{2 m} -
U(z_1)\equiv i\frac{\partial}{\partial t_1} + \frac{\Delta_1^{\perp}}{2 m} +
\hat{L}(z_1), 
\end{eqnarray*}
where $\Delta_1^{\perp}\equiv \partial^2_{x_1} + \partial^2_{y_1}$.
Since we are interested in the steady state solution,
one can see that, the properties of this operator can be fully
accounted for if we 
introduce the function $g(z,k)$, such that
\begin{equation}
\left[\frac{1}{2m} \frac{\partial^2}{\partial z^2} -
U(z) \right] 
g(z,k)=-\epsilon_k g(z,k),
\end{equation}
where $\epsilon_k=k^2/2m$.
Since the operator $\hat{L}$ is Hermitian, we will use the
property of its eigen-functions $g(z,k)$:
\begin{equation}
\int_{-\infty}^{\infty} g(z,k_1) g^*(z,k_2) dz= 2 \pi \delta(k_1 - k_2)
\label{orthogonality} 
\end{equation}

Now let us solve for $G^{(0)-+}_{12}$, which satisfies
\begin{equation}
\hat{G}^{-1}_1 G^{(0)-+}_{12}=0. \label{kinetic-zero}
\end{equation}
It is easier to find $G^{(0)-+}_{12}$ by using its definition
\begin{equation}
G^{(0)-+}_{12}=i\left<\hat{\psi}_2^{\dagger} \hat{\psi}_1 \right>,
\label{mp-definition} 
\end{equation}
where $\hat{\psi}^{\dagger}, \hat{\psi}$ are the Heisenberg operators of
creation and annihilation of the particles at a point $X$, and $<..>$
means quantum and statistical averaging. 
Since $\hat{\psi}$ evolves according to $\hat{G}^{-1} \hat{\psi}=0$,
one can see, that it is equal to
\begin{equation}
\hat{\psi}(X)=\sum_{{\bf q}_{\perp} k} \hat{a}_{{\bf q}_{\perp} k}
g(z,k) e^{-i 
\epsilon_{{\bf q}_{\perp} k} t + i {\bf q}_{\perp}\cdot {\bf
x}^{\perp}} \label{psi}
\end{equation}
Here $\hat{a}_{{\bf q}_{\perp} k}$ is the annihilation operator of the
particle with momentum $({\bf q}_{\perp},k)$, and
$\epsilon_{{\bf q}_{\perp} k}\equiv {\bf q}_{\perp}^2/2m + k^2/2m$.

We substitute $\psi(X)$ from (\ref{psi}) into
Eq.~(\ref{mp-definition}) and obtain
\begin{eqnarray*}
G^{(0)-+}_{\om {\bf q}_{\perp}}(z_1,z_2)=2 \pi i
\int_{-\infty}^{\infty} \frac{dk}{2 \pi}&& \hspace*{15mm}
\end{eqnarray*}
\begin{equation}
 \delta(\om - \epsilon_{{\bf q}_{\perp} k} + \mu) n({\bf q}_{\perp} k)
g(z_1,k) g^*(z_2,k), \label{mp-zero}
\end{equation}
where we denote $\left<a^{\dagger}_{{\bf q}_{\perp} k} a_{{\bf
q}_{\perp} k}\right>\equiv n({\bf q}_{\perp} k)$; $\mu$ is the
chemical potential.

It is easy to see that, in thermodynamic equilibrium we are
considering, the function $n({\bf q}_{\perp} k)$ is equal to Fermi
distribution $n_F(\epsilon_{{\bf q}_{\perp} k})$. 
To convince ourselves we take
$G^{(0)-+}_{\om {\bf q}_{\perp}}(z_1,z_2)$ at such values of
$z_1,z_2$, when $U(z)$ is negligible and $g(z,k)\Rightarrow
exp(ikz)$. One more Fourier transform $\int
exp[-iq(z_1 - z_2)] d(z_1 - z_2)$ will give $G^{(0)-+}_{\om {\bf q}}$,
which has to coincide with the known expression~\cite{landau-10}
$G^{(0)-+}_{\om {\bf q}}=2 \pi i \delta(\om - \epsilon_{\bf
q} + \mu) n_F(\epsilon_{\bf q})$.

Before proceeding further with solving for $G^{-+}(X_1,X_2)$ we first
find $K^{(0)}$, which is the tunneling rate obtained by using
$G^{(0)-+}_{12}$ in Eq.~(\ref{rate-general}). 
Since we are interested in $K^{(0)}$ behind the barrier, where $U(z)=0$, we
can exchange $\partial_{t_1}$ for $i\Delta_1/2m$ as is clear from the
form of the operator $\hat{G}^{-1}_1$ and Eq.~(\ref{kinetic-zero}).
Then the formula for the rate
$K^{(0)}$ takes the form:
\begin{equation}
K^{(0)}(z)=\frac{i}{2m}\int \frac{d\om}{2 \pi} \int\frac{d^2{\bf
q_{\perp}}}{(2 \pi)^2} \left[\frac{\partial^2}{\partial \zeta^2}
G^{(0)-+}_{\om {\bf q}_{\perp}}(z,\zeta)\right]_{\zeta=0}, \label{rate-zero}
\end{equation}
where $z=(z_1 + z_2)/2$, $\zeta=z_1 - z_2$.

In deriving (\ref{rate-zero}) we expressed $G^{(0)-+}_{12}$ through
its Fourier transform $G^{(0)-+}_{\om {\bf q}_{\perp}}(z,\zeta)$ with
respect to ``fast'' space ${\bf r}_{\perp}={\bf r}^{\perp}_1 - {\bf
r}^{\perp}_2$ and time $\tau=t_1 - t_2$
variables and took the limit ${\bf r},\tau \rightarrow 0$ as
prescribed by Eq.~(\ref{rate-general}).

Finally we need to know the function $g(z,k)$ in the region behind the
barrier. We will use the standard quasiclassical
expression~\cite{landau-3}
\begin{equation}
g(z,k)=\frac{C_k}{\tilde{q}(z,k)^{1/2}} exp\left[i\int_{z_*}^z
\tilde{q}(z',k) dz' \right] \label{g-quasicl} 
\end{equation}
\begin{equation}
C_k=A_k exp\left[-\int_{z_a}^{z_b(k)}
\tilde{q}(z',k) dz'\right]  \label{ck}
\end{equation}
\begin{equation}
\tilde{q}(z,k)=\sqrt{2m(\epsilon_k - U(z))} \label{qk}
\end{equation}
The square of the exponential factor in Eq.~(\ref{ck}) is the
tunneling coefficient, $\tilde{W}(k)$~\cite{landau-4}.
We multiply it by a factor $A_k=(S(\epsilon_k)/\epsilon_k)^{1/2}$, which
takes into account nuclear 
physics effects, not considered here. $S(\epsilon_k)$ is the astrophysical
factor~\cite{fusion-adelberger}. 

To find $K^{(0)}$ from Eq.~(\ref{rate-zero}) we substitute
Eqs.~(\ref{g-quasicl})-(\ref{qk}) in Eq.~(\ref{mp-zero}), use
$z_1=z+\zeta/2$, 
$z_2=z-\zeta/2$ and take
$\partial_{\zeta}^2$ derivative in the quasiclassical sense keeping
only zero-th order terms. The result of the differentiation is
\begin{equation}
\left[\frac{\partial^2}{\partial \zeta^2} \left[g(z_1,k) g^*(z_2,k)\right] \right]_{\zeta=0}=-\tilde{q}(z,k) C_k^2
\label{derivative} 
\end{equation}
Note, that $\tilde{q}(z,k)=k$ is
true behind the barrier. 

The problem of tunneling collision between particles $m_1$ and $m_2$
can be reduced to the one-dimensional motion of a particle with
reduced mass
$m_r=m_1 m_2/(m_1 + m_2)$ in the external
potentianl $U(r)$~\cite{landau-3}. We make this reduction in the original
Hamiltonian and then second
quantize it. Then we use spherical coordinates, assume spherical
symmetry of the tunneling collision and
perform
calculations similar to those explained above.
The final result for $K^{(0)}$ is
\begin{equation}
K^{(0)}=n_0 \int_{0}^{\infty}
\frac{4 \pi k^2 dk}{(2 \pi)^3} \bar{n}_M(k) \frac{k}{m_r}
\frac{S(\bar{\epsilon}_k)}{\bar{\epsilon}_k} W(k), \label{rate-zero-final}
\end{equation}
where $\bar{\epsilon}_k=k^2/2m_r$. $W(k)$ is obtained from
$\tilde{W}(k)$ by $z\rightarrow r$, $\bar{n}_M(k)$ is the Maxwell
distribution which depends on $m_r$.

It agrees with the 
answer obtained in~\cite{rate-gamow} by averaging the quasiclassical
tunneling factor $W(k)$ over the Maxwell distribution,
$\left<n_M(v)\sigma(m_rv^2/2)v\right>$. Hence, we see, that the
dispersion relation $\delta(\om - \epsilon_{\bf q})$ was implicitly
used in \cite{rate-gamow}, which corresponds to an approximation of
instantaneous, two-body collisions. Now we will not make this 
assumption and proceed with finding the 
tunneling rate, $K$, of the particles, which always collide and hence are
never ``in between collisions''. 

By using~\cite{landau-10,statistical-mechanics-kadanov}, we find 
\begin{eqnarray*}
\left(\frac{q}{m}\frac{\partial}{\partial z}  +  U'(z)
\frac{\partial}{\partial q}\right) G^{-+}_{\om {\bf
q}}(z)=- \gamma_{\om {\bf q}}(z) G^{-+}_{\om {\bf q}}(z)
\end{eqnarray*}
\begin{equation}
-\gamma_{\om {\bf
q}}(z) n_F(\om) \left(G^R_{\om {\bf q}}(z) - G^A_{\om {\bf
q}}(z)\right), \label{kinetic-inter}
\end{equation}
where we denote $\Sigma^{R} - \Sigma^{A}\equiv i \gamma_{\om
{\bf q}}(z)\equiv i \gamma_{\om {\bf q}_{\perp}}(q,z)$.
Note, that $\gamma_{\om {\bf q}}(z)$ is a non-linear function of
$G^{\alpha \beta}$~\cite{statistical-mechanics-kadanov,landau-10}.    
We now analyze Eq.~(\ref{kinetic-inter}) qualitatively,
which will help us to find that part of the
solution, which makes the largest contribution to $K$.

In the region {\em away} from the barrier we can neglect the
LHS of Eq.~(\ref{kinetic-inter}), and obtain:
\begin{equation}
\tilde{G}^{-+}_{\om {\bf q}}(z)=-n_F(\om)\left(G^R_{\om {\bf q}}(z) -
G^A_{\om {\bf q}}(z)\right) \label{mp-tilde}
\end{equation} 
Therefore, Eq.~(\ref{kinetic-inter}) will
allow us to propagate this solution into the region {\em behind} the
barrier.

We can find the first integral and formally integrate
Eq.~(\ref{kinetic-inter}) along trajectories, 
which will lead to a sum of solutions of homogeneous and
"inhomogeneous" equations (due to the last term in
(\ref{kinetic-inter})). As can be seen from Eq.~(\ref{kinetic-inter}),
"inhomogeneous" solution will involve the integral of the product of
$\gamma_{\om {\bf q}}(z)$ and $G^R_{\om {\bf q}}(z)$ or $G^{R*}=G^A$. 
Qualitatively,
it means, that in the region of $z$ behind the barrier it will be
proportional to a product of at least two exponential 
factors $W(k)$. This is so, because both $\gamma_{\om
{\bf q}}(z)$ and $G^R_{\om {\bf q}}(z)$ will depend on the integral of
the product $g(z_1,k) g^*(z_2,k)$ and such product is $\propto W(k)$
as can be seen from Eqs.~(\ref{g-quasicl})-(\ref{qk}).

It is obvious that, the solution of the "homogeneous" equation,
$G^{-+}_f$, involves only one factor of $W(k)$, since $G^{-+}_f
\propto g g^*$, see Eq.~(\ref{kinetic-inter}). Therefore, we will find
only the solution to the "homogeneous" equation, which we will use to
obtain $K$. We supplement this homogeneous equation with the boundary
condition, $\tilde{G}^{-+}$ from Eq.~(\ref{mp-tilde}) imposed at
$z=z_*$ away from the barrier, see Fig.~1. 

Since the largest contribution to $K$ is made by $G^{-+}_f$ one can
see, that we can use Eq.~(\ref{rate-zero}) to find $K$ if we substitute
$G^{-+}_f$ instead of $G^{(0)-+}$ in (\ref{rate-zero}).
Therefore, we find $G^{-+}_f$ with
dependence on $z_1, z_2$ from the very begining by 
applying the transformation
\begin{equation}
\int \int g(z_1,k_1) g^*(z_2,k_2) \frac{d k_1}{2 \pi} \frac{d k_2}{2
\pi} \label{transformation}
\end{equation}
to Eq.~(\ref{kinetic-general}). We use the boundary condition,
Eq.~(\ref{mp-tilde}) and obtain the
following answer: 
\begin{equation}
G^{-+}_f=2 \pi i \, n_F(\om) \int g_k(z_1) \delta_{\gamma}(\om -
\epsilon_{{\bf q}_{\perp}k}) g^*_k(z_2) \frac{dk}{2 \pi} \label{mp-final}
\end{equation}
where $g_k(z)\equiv g(z,k)$ and we used the well-known equilibrium
solution~\cite{statistical-mechanics-kadanov} 
\begin{eqnarray*}
G^R_{\om {\bf q}_{\perp}}(k) - G^A_{\om {\bf
q}_{\perp}}(k)=\frac{\gamma_{\om {\bf q}_{\perp}}(k)}{(\om -
\epsilon_{{\bf q}_{\perp}k})^2 + (\gamma_{\om {\bf q}_{\perp}}(k))^2}
\end{eqnarray*}
\begin{equation}
G^R_{\om {\bf q}_{\perp}}(k) - G^A_{\om {\bf
q}_{\perp}}(k)\equiv \delta_{\gamma}(\om - \epsilon_{{\bf q}_{\perp}k})
\label{r-equil} 
\end{equation}

Comparing Eq.~(\ref{mp-zero}) with (\ref{mp-final}),
(\ref{r-equil}) we see, 
that the effect of collisions, which makes the largest contribution to
$K$ can be described as $\delta(\om - \epsilon_{{\bf q}_{\perp}k})
\Rightarrow \delta_{\gamma}(\om - \epsilon_{{\bf q}_{\perp}k})$, as can
be expected on the intuitive grounds. Note also, that we can obtain
the answer (\ref{mp-final}), (\ref{r-equil}) by following the same
steps leading to Eq.~(\ref{mp-zero}) if we make a substitution
$\epsilon_{{\bf q}_{\perp} k} \rightarrow \epsilon_{{\bf q}_{\perp} k}
+ i \gamma_{{\bf q}_{\perp} k}$. This means that, we have to use
{\em damped} oscillators instead of {\em undamped} ones for second
quantization of {\em colliding} particles.

Now we substitute Eqs.~(\ref{mp-final}), (\ref{r-equil}) into 
(\ref{rate-zero}) and integrate $\int d\om/2 \pi$ by using
the result of~\cite{tails-galitski}:
\begin{equation}
\int \frac{d \om}{2 \pi} n_M(\om) \delta_{\gamma}(\om - \epsilon_{{\bf
q}_{\perp} k})=n_M(\epsilon_{{\bf q}_{\perp} k}) + \delta
n_{\gamma}({\bf q}_{\perp}k), \label{tail}
\end{equation}
where $\delta n_{\gamma}({\bf q}_{\perp}k)$ is the ``quantum tail''. 

At this point we are ready to perform the same steps as those leading
to Eq.~(\ref{rate-zero-final}) in order to describe the tunneling of
particles $m_1$ and $m_2$. The answer for $K$ is: 
\begin{eqnarray*}
K=n_0 \int_0^{\infty} \frac{4 \pi p^2 dp}{(2 \pi)^3} \frac{S(\bar{\epsilon}_p)}{\bar{\epsilon}_p}&& \hspace*{25mm}
\end{eqnarray*}
\begin{equation}
 \frac{p}{m_r} W(p) \left[\bar{n}_M(p) +
\delta \bar{n}_{\gamma} (p) \right]\equiv K_M + K_{\gamma}. \label{k-semifinal}
\end{equation}

We now evaluate $\Sigma^R$ through collision frequency, $\nu_T$, as in
\cite{tails-galitski,sigma-galitski}, carry out the integral in
(\ref{k-semifinal}) and form the ratio $r_{12}=K_{\gamma}/K_M$:
\begin{equation}
r_{12}=\frac{3^{19/2}}{8 \pi^{3/2}} \sum_{j} \frac{h
\nu_T(m_{coll})}{T} \left(\frac{m_{coll}}{m_r}\right)^{7/2}
\frac{e^{\tau_{12}}}{\tau_{12}^8} \label{k-final}
\end{equation}
Here $m_{coll}=m_r m_j/(m_r + m_j)$, $m_r=m_1 m_2/(m_1 + m_2)$,
$\tau_{12}=3(\pi/2)^{2/3} 
\left(\frac{100 Z_1^2 Z_2^2 A_{12}}{T_k}\right)^{1/3}$, $A_{12}=A_1
A_2/(A_1 + A_2)$, $T_k$ is the temperature in keV; $A_1$, $A_2$ are
the atomic numbers of tunneling particles, 
$m_j$ is the mass of the
background particles colliding with tunneling particles. Summation is
over all species of colliding background particles, including
tunneling species.

This result, (\ref{k-semifinal}), (\ref{k-final}), supports the idea
of averaging of $\sigma v$ over momentum 
distribution postulated in ~\cite{neutrino-vlad}, but differs from the
answer of ~\cite{neutrino-vlad} by a factor $(m_{coll}/m_r)^3$ under the
summation sign.

Consider, for example, a $DT$ plasma of density
$\rho=10\, {\rm g/cm}^2$ and temperature $T=0.1$~keV. Then
Eq.~(\ref{k-final}) will lead to the rate accelerated by five orders
of magnitude as compared to the conventional answer. We find the ratio
$r_{DT}=K^{\gamma}_{DT}/K^M_{DT}=8 \cdot 10^4$, with
$K^M_{DT}=2.7 \cdot 10^{-30} {\rm cm}^3/{\rm sec}$,
$K^{\gamma}_{DT}=2.1 \cdot 10^{-25} {\rm cm}^3/{\rm sec}$.  
In case of heavier elements and not very small $h \nu_T/T$, ratio $r_{ij}$
becomes much higher~\cite{neutrino-vlad}. 

I would like to acknowledge helpful discussions with N.~J.~Fisch.
This work was supported by NSF--DOE grant
DE-FG02-97ER54436.

\end{multicols}

\begin{thebibliography}{10}

\bibitem{tails-galitski}
V.~M. Galitski and V.~V. Yakimets, ZhETF {\bf 51},  957  (1966).

\bibitem{tails-cross-section-star}
A.~N. Starostin and N.~L. Aleksandrov, Phys.~Plasmas {\bf 5}, 2127
(1998); N.~L. Aleksandrov and A.~N. Starostin,ZhETF {\bf 113},  1661  (1998).

\bibitem{neutrino-vlad}
A.~N.~Starostin, V.~I.~Savchenko, N.~J.~Fisch submitted to Phys.~Rev.~Lett.

\bibitem{statistical-mechanics-kadanov}
L.~P. Kadanoff and G. Baym, {\em Quantum Statistical Mechanics}
  (W.~A.~Benjamin, New York, 1962).

\bibitem{rate-gamow}
G. Gamow and E. Teller, Phys.~Rev. {\bf 53},  608  (1938).

\bibitem{landau-10}
E.~M. Lifshitz and L.~P. Pitaevskii, {\em Physical Kinetics} (Pergamon Press,
  New York, 1977).

\bibitem{diagramms-keldish}
L.~V. Keldish, Sov. Phys. JETP {\bf 20},  1018  (1965).

\bibitem{diagramms-korenman}
V. Korenman, Ann. Phys. {\bf 39},  72  (1966).

\bibitem{landau-3}
L.~D. Landau and E.~M. Lifshitz, {\em Quantum Mechanics} (Pergamon Press, New
  York, 1977).

\bibitem{landau-4}
E.~M. Lifshitz and L.~P. Pitaevskii, {\em Relativistic Quantum Theory}
  (Pergamon Press, New York, 1977).

\bibitem{fusion-adelberger}
E.~G. Adelberger {\it et~al.}, Rev. Mod. Phys. {\bf 70},  1265  (1998).

\bibitem{sigma-galitski}
V.~M. Galitski, JETP {\bf 34},  104  (1958).

\end{thebibliography}
\end{document}